\documentclass[pra,showpacs,showkeys,amssymb,amsmath]{revtex4}
\usepackage{amssymb}
\usepackage{amsmath}
\usepackage{graphicx}
\usepackage{color}
\usepackage{slashed}
\newcommand{\be}{\begin{equation}}
\newcommand{\ee}{\end{equation}}
\newcommand{\ben}{\begin{eqnarray}}
\newcommand{\een}{\end{eqnarray}}

\begin{document}
\title{A Model to Study Finite-Size and Magnetic Effects on the Phase
Transition of a Fermion Interacting System}
%%%%%%%%%%%%%%%%%%%%%%%%%%%%%%%%%%%%%
%%%%%%%%%%%%%%%%%%%%%%%%%%%%%%%%%%%%%%%%%%
\author{Emerson B. S. Corr\^ea}
\email[]{emersoncbpf@gmail.com}
\affiliation{Centro Brasileiro de Pesquisas F\'{\i}sicas/MCTI,
	22290-180, Rio de Janeiro, RJ, Brazil}
\affiliation{Faculdade de F\'isica, Universidade Federal do Sul e Sudeste do Par\'a, 68505-080, Marab\'a, PA, Brazil}
\author{C\'{e}sar A. Linhares}
\email[]{linharescesar@gmail.com}
\affiliation{Instituto de F\'{\i}sica, Universidade do Estado do Rio
de Janeiro, 20559-900, Rio de Janeiro, RJ, Brazil}
\author{Adolfo P. C. Malbouisson}
\email[]{adolfo@cbpf.br}
\affiliation{Centro Brasileiro de Pesquisas F\'{\i}sicas/MCTI,
22290-180, Rio de Janeiro, RJ, Brazil}
%%%%%%%%%%%%%%%%%%%%%%%%%%%%%%%%%%%%%%%%%%%%%%%%%%%%%%%%%%%%%
%%%%%%%%%%%%%%%%%%%%%%%%%%%%%%%%%%%%%%%%%%%%%%%%%%%%%%%%%%%%%
\begin{abstract}
We present a model to study effects from an external magnetic field, chemical potential, and finite size, on the phase structure of a massive four- and six-fermion interacting system. These effects are introduced by a method of compactification of coordinates, a generalization of the standard Matsubara prescription. Through the compactification of the $z$ coordinate and of imaginary time, we describe a heated system with the shape of a film of
thickness $L$, at temperature $\beta^{-1}$ undergoing first- or second-order phase transition. We have found a strong dependence of the temperature transition on the constants couplings $\lambda$ and $\eta$. Besides magnetic catalysis and symmetry breaking for both kinds of transition, we have found an inverse symmetry breaking phenomenon with respect to first-order phase transition.
\end{abstract}
\keywords{phase transitions; four- and six-fermion interaction; finite-size effects}
\pacs{11.30.Qc; 11.10.Wx; 11.10.Kk}
\maketitle

\section{Introduction}
Quantum Chromodynamics (QCD) is currently accepted as the theory for describing hadronic matter;  its Lagrangian is assumed to give all the dynamics of the strong interactions. However, it is not easy to get analytical results from it; in particular, when one takes into account temperature and density effects. This is due to the mathematically intricate structure of QCD,  and for this reason simpler, effective models have called attention over the last few decades, for instance, the Nambu--Jona-Lasinio (NJL)\cite{NJL,NJL1}, Gross--Neveu (GN)\cite{gn1} and MIT Bag \cite{Chodos} models. 
In particular, the GN model shares some interesting  features with QCD, like asymptotic freedom and dynamical symmetry breaking; moreover, it has a peculiar simplicity from the mathematical point of view. Extensions of this effective model have been made in many articles with applications to both condensed matter and  hadronic physics~\cite
{gn2,gn5,gn7,gn8,gn9,gn11,gn12,gn13,gn14,gn18,gn19,PLB04,GNL,AMM2,AMM3}.

Interesting effects on the phase transitions in these effective models can be investigated. For example, their behaviour in a uniform magnetic background. This is a very important effect. In  ultrarelativistic heavy-ion collisions  strong magnetic fields ($B \approx 10^{18} \ G$) can be created perpendicularly to the collision plane. Also, weaker magnetic fields  ($B \approx 10^{15} \ G$) in neutron stars could have an influence in the phase transition of dense hadronic matter \cite{livromag}. An investigation of the first- and second-order phase transitions in a magnetic background of an extended  GN-like model, including four- and six-fermion interactions, has not yet been done, up to our knowledge. We intend to perform this work here.

Another important factor of influence on the phase transition is size restriction. The methodology to deal with this effect was improved in Refs. \cite{Birrel,livro,AOP11,review} inspired on the imaginary-time formalism, the so-called Matsubara formalism \cite{Matsubara}. Before presenting size restriction on the model as a quantum field theory on a toroidal topology, let us briefly recall the imaginary-time formalism. It is constructed in the following way: in a $(1+3)$-dimensional Euclidean manifold, for example, the coordinates will be called $(\tau,x,y,z)$. To introduce temperature on the system, the Matsubara prescription tells us to compactify the $\tau$ coordinate through the imposition of periodic or antiperiodic boundary conditions. The imaginary-time formalism defines a theory on a space with topology $\Gamma^{1}_{4} = S^{1}\times {\mathbb{R}}^{3}$, where $S^1$  is a circumference with length $\beta = {1/k_{B}T}$. Here we consider a generalization in which the Matsubara prescription is extended to a spatial dimension. Specifically, we will compactify the system in both $\tau$ and $z$ directions ($z$ being the same direction of the applied magnetic field), with a compactification in the $z$ direction of length $L$; the topology of the system becomes $\Gamma_{4}^{2} = (S^{1})^{2}\times {\mathbb{R}}^{4-2}$. Then, as argued in \cite{livro,review}, $L$ can be interpreted as the separation between two parallel planes orthogonal to the $z$ direction. Thus, after performing these two compactifications, we get a system in thermal equilibrium in the form of a film. Notice that this is a particular case of a general field theory defined on a toroidal topology\cite{review}.  

In this paper we explore the effects coming from finite size,  chemical potential and from a magnetic background on a  massive version of an extended fermion interacting model  for a system undergoing a first- and a second-order phase transitions. As in~\cite{EPL12}, we deal with a one-component fermionic field and we would like to investigate how the transition temperature is affected by spatial boundaries and the magnetic field. Indeed, this is a generalization, to include a magnetic background, of the work developed in \cite{paperfermionico}.

The paper is organized as follows. In section II we present the model and fix the conventions on the parameters to ensure the system stability and also the Ritus propagator for a fermion in a magnetic background. In  section III, we calculate the two Feynman diagrams used to perform corrections due to the temperature, chemical potential and size restriction. In section IV we perform an analysis for for both first- and second-order phase transitions.  Indeed, we show that the catalysis phenomenon occurs for a second-order phase transition, independently of the value of the coupling constant as well as symmetry breaking at temperatures lower than the critical one. However, for the first-order phase transition, we find a rather non intuitive phenomenon, namely, {\it inverse symmetry breaking}: there are two different critical temperatures. Below the smaller critical temperature, $t_{c1}$, we have the usual broken-symmetry phase. Above the higher critical temperature, $t_{c2}$, we have also found a broken phase. For intermediate temperatures, $t_{c1} < t < t_{c2}$, we find a symmetric phase. This phenomenon of an {\it{inverse symmetry breaking}} has been already mentioned in the literature \cite{Bimonte,Ramos1,Hoa,Sakamoto}.
In  section V we present our conclusions and notice that no transition exists below a minimal size of the system in both first- and second-order phase transition and that density (chemical potential) has little influence on the phase structure. We will use natural units $c=\hbar=k_B = 1$.
\section{The Model}
We start by considering the Hamiltonian density in Euclidean space for a generalization of the massive Gross--Neveu model which includes quartic as well as sextic fermionic interactions,
\begin{eqnarray}
\mathcal{H} = \bar{\psi}(x)(i\slashed{D}-m_{0}^{\prime})\psi (x)-\frac{\lambda _{0}}{2}\left[ \bar{\psi}(x)\psi (x)\right]
^{2}+\frac{\eta_{0}}{3}\left[\bar{\psi}(x)\psi (x)\right]
^{3},
\label{GN}
\end{eqnarray}
where $x = (\tau,x,y,z \hspace{0.05cm} )$ and we use the  chiral representation for the Dirac matrices. The covariant derivative is given by $D_{\mu} =\ \partial_{\mu} + ie A_{\mu}$, where $A_{\mu}=(0,0,Bx,0)$. This gauge choice makes ${\bf{B}} = B\hat{z}$, $B$ being a constant \cite{Lawrie}. The parameter $m_{0}^{\prime}$ can be chosen as $\pm m_{0}$ (the sign is important to ensure the stability of the system). This model is useful to study both first- and second-order phase transitions. To describe a first-order transition, we choose $m_{0}^{\prime} = - m_0$. On the other hand, to study a second-order transition, we fix  $m_{0}^{\prime} = + m_0$. 

We assume a free-energy density of the  Ginzburg--Landau  type 
\begin{eqnarray}
\mathcal{F}=\mathcal{A}\phi ^{2}(x)+\mathcal{B}\phi^{4}(x)+\mathcal{C}\phi^{6}(x),
\end{eqnarray}
where $\phi (x)$ denotes the order parameter of the model, namely, $\phi (x)=\sqrt{\left\langle \bar{\psi}(x)\psi
(x)\right\rangle_{\beta} }$ where, $\left\langle \cdot \right\rangle_{\beta} $ means a thermal average in the grand-canonical ensemble.
We consider $\mathcal{A}=-m, \  \mathcal{B}=-\lambda_{0}/2, \  \mathcal{C}=\eta_{0}/3$, where $m = m(T,\mu,L,B)$ includes temperature, chemical potential, finite size and magnetic field corrections to the $m_{0}^{\prime}$ parameter, defined at zero temperature. We do not compute corrections to the coupling constants.

The Euclidean Feynman propagator is calculated via the Ritus method, \cite{Ritus,Mexicanos},
\begin{equation}
\mathcal{S}(A) = \frac{\omega }{2\pi }\sum_{s=\pm 1}^{}\sum_{\ell =0}^{\infty }\int\frac{dp_\tau}{2\pi}\frac{dp_z}{2\pi} \frac{(%
\slashed{\bar{p}}-m_{0}^{\prime})}{\bar{p}^{2}+m_{0}^{2}},
\end{equation}
where ${\overline{p}}^{2} =p_{\tau}^{2}+p_{z}^{2}+\overline{p}_{1}^{2}+\overline{p}_{2}^{2}$ ; $\overline{p}_{1}^{2}+\overline{p}_{2}^{2} =\omega (2\ell +1- s ),$  $s=\pm 1$ being the spin variable and $\omega \equiv eB$ the cyclotron frequency. The natural numbers $\ell$ label the Landau levels.
\section{Corrections to the Mass Parameter $\large{m_{0}^{\prime}}$}
To obtain corrections due to temperature $\beta$, finite size $L$, chemical potential $\mu$ and external field $\omega$, we use just two Feynman diagrams, namely, the {\textit{Tadpole}} e the {\textit{Shoestring}}, depicted in Fig.~$1$:
\begin{center}
\includegraphics[scale=0.2]{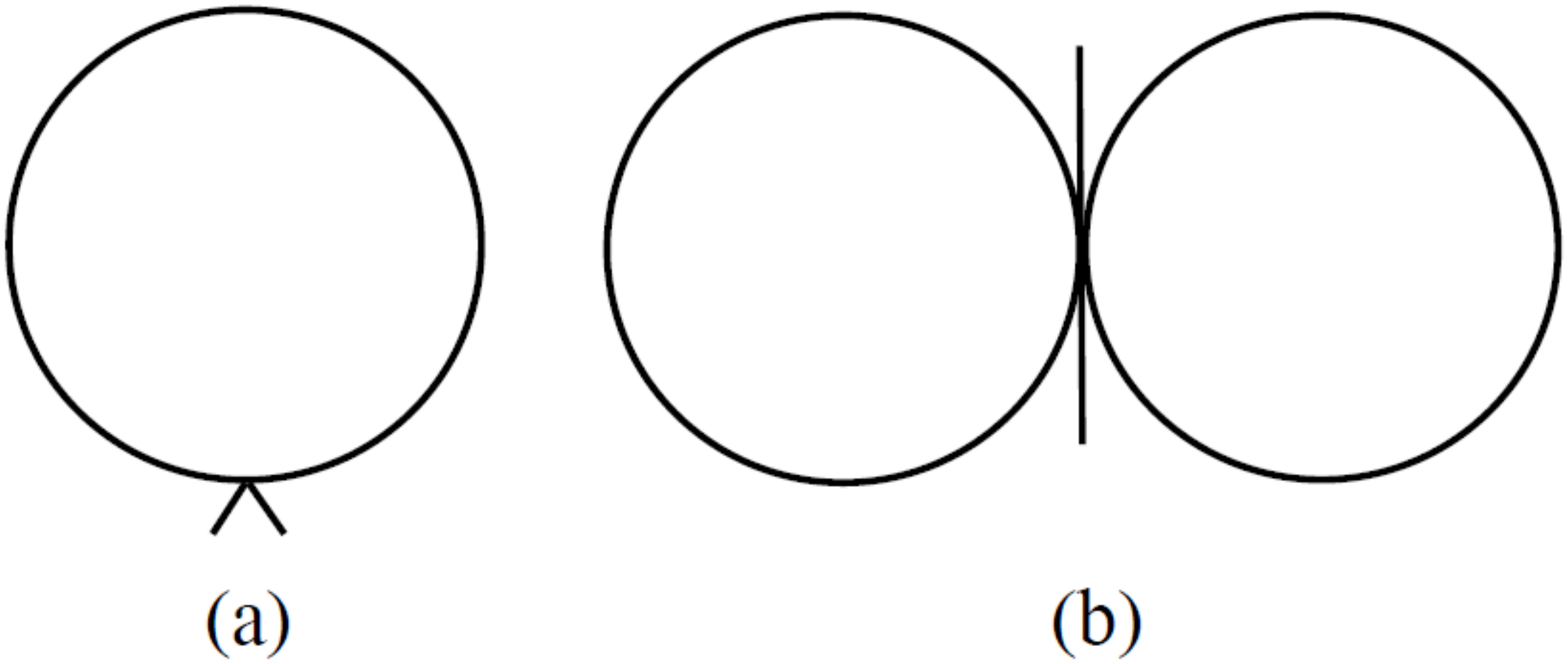}
\label{Fig1}
\end{center}
{Fig.~1 - \it{Feynman diagrams to be considered: (a) Tadpole; (b) Shoestring.}}

The self-energy becomes
\begin{eqnarray}
m(\beta ,L,\mu ,\omega )=m_{0}^{\prime}+\Sigma^{(a)} (\beta ,L,\mu ,\omega )+\Sigma^{(b)} (\beta ,L,\mu ,\omega ),
\label{cri1}
\end{eqnarray}
where $\Sigma$ represents the corresponding diagram to be calculated.
\subsection{Tadpole diagram contribution}
Following the method developed in \cite{livro}, we start from
\begin{eqnarray*}
\Sigma^{(a)} (\beta ,L,\mu ,\omega )&=& \lambda_{0}{\rm{tr}}\left[
\frac{\omega }{2\pi }\sum_{s=\pm 1} \sum_{\ell =0}^{\infty }\int\frac{dp_\tau}{2\pi}\frac{dp_z}{2\pi} \frac{(%
\slashed{\bar{p}}-m_{0}^{\prime})}{\bar{p}^{2}+m_{0}^{2}}\right].
\end{eqnarray*}
To insert temperature, chemical potential and finite size, we use the method described in Ref.\cite{livro}. In this way, we introduce two compactifications:
\begin{eqnarray*}
&&\int \frac{d{p}_{\tau}}{2\pi } \rightarrow \frac{1}{\beta }
\sum_{n_{1}=-\infty }^{+\infty } \ ; \  {p}_{\tau}\rightarrow \frac{2\pi }{
\beta }\left( n_{1}+\frac{1}{2}-i\frac{\mu \beta }{2\pi }\right); \  \nonumber \\  
&&\int \frac{d{p}_{z}}{2\pi } \rightarrow \frac{1}{L}
\sum_{n_{2}=-\infty }^{+\infty } \ ; \ {p}_{z}\rightarrow \frac{2\pi }{L%
}\left( n_{2}+\frac{1}{2}\right).  \nonumber \\
\label{Matsubara} 
\end{eqnarray*}
Therefore, we obtain
\begin{eqnarray}
\Sigma^{(a)} (\beta ,L,\mu ,\omega ) &=& \lambda_{0}\left[\frac{\omega }{2\pi }\sum_{s=\pm 1}
\sum_{\ell =0}^{\infty } \ \sum_{n_1,n_2 = -\infty}^{+\infty}\frac{-4m_{0}^{\prime}}{\beta L 4\pi^2 m_{0}^{2}} \right. \nonumber \\
&&\left. \times \left(\frac{1}{a_1(n_1 - b_1)^{2}+a_2(n_2 - b_2)^{2}+c_{\ell,s}^{2}}\right)\right],
\label{eq5}
\end{eqnarray}
where we have defined the dimensionless quantities 
\begin{eqnarray*}
&&a_{1}=(m_{0}\beta )^{-2}=t^{2} \ ; \ a_{2}=(m_{0}L)^{-2}=\xi ^{2} \frac{}{}; 
\nonumber \\ 
&& c_{\ell,s}^{2} = [\delta(2\ell+1-s)+1]/4\pi^2 \ ; \ \gamma =\mu /m_{0} \ ; \ \delta = \omega/m_{0}^{2} \frac{}{}  ; \nonumber \\ 
&& \ b_{1}=i\beta \mu /2\pi -1/2 = i\gamma /t2\pi -1/2 \ ; \ b_{2}=-{1/2}.
\label{reduzidos}
\end{eqnarray*}%
In terms of these dimensionless quantities, we can rewrite the expression in Eq. (\ref{eq5}) as
\begin{eqnarray}
\Sigma^{(a)} (t,\xi,\gamma,\delta )&=&\lambda_{0}\left\{-\frac{\delta}{2\pi^3}\left(m_{0}^{2}m_{0}^{\prime}\sqrt{a_1a_2}\right) Z_{2}^{c_{\ell,s}^{2}}\left[1,\{a_j\},\{b_j\}\right]\right\},
\label{tad1}
\end{eqnarray}
where we have introduced the Epstein--Hurwitz inhomogeneous zeta function,
\begin{eqnarray*}
Z_{2}^{c_{\ell,s}^{2}}\left[\nu,\{a_j\},\{b_j\}\right] &=& \sum_{n_{1},n_2=-\infty }^{+\infty } \left[\frac{}{}a_1(n_1-b_1)^{2}+a_2(n_2-b_2)^{2}+c_{\ell,s}^{2}\right]^{-\nu},
\end{eqnarray*}
with $j=1,2$. Its analytical extension to the whole complex set of values on the $\nu-plane$ is given by\cite{ep2}
\begin{eqnarray}
Z_{2}^{c_{\ell,s}^{2}}\left[\nu ,\{a_j\},\{b_j\}\right]  &=& \frac{\pi(c_{\ell,s}^{2})^{1-\nu}}{\sqrt{a_1a_2}}\frac{\Gamma(\nu-1)}{\Gamma(\nu)}+\frac{4\pi^{\nu}}{\sqrt{a_1a_2}}\frac{1}{\Gamma(\nu)}{\widetilde{Z}}_{2}^{c_{\ell,s}^{2}}(\nu ),
\label{Zeta2} 
\end{eqnarray}
where
\begin{eqnarray}
{\widetilde{Z}}_{2}^{c_{\ell,s}^{2}}(\nu)&=&\sum_{j=1}^{2}\sum_{n_{j}=1}^{\infty }\cos(2\pi n_j b_j) \left(\frac{n_j}{c_{\ell,s}\sqrt{a_j}}\right)^{\nu-1} K_{\nu -1}\left(\frac{2\pi c_{\ell,s} n_j}{\sqrt{a_j}}\right)\nonumber \\
&&+2\sum_{n_{1},n_{2}=1}^{\infty }\cos(2\pi n_1 b_1)\cos(2\pi n_2 b_2) \left(\frac{1}{c_{\ell,s}}\sqrt{\frac{n_1^2}{a_1}+\frac{n_2^{2}}{a_2}}\right)^{\nu-1} \nonumber \\
&&\times K_{\nu -1}\left(2\pi c_{\ell,s}\sqrt{\frac{n_1^2}{a_1}+\frac{n_2^{2}}{a_2}}\right).
\label{Zeta3}
\end{eqnarray}
Since $\Gamma(\nu-1)/\Gamma(\nu) = 1/(\nu-1)$, we  see that the first term in Eq.(\ref{Zeta2}) is  divergent for $\nu=1$. We apply the modified minimal subtraction prescription to eliminate this term, so that Eq. (\ref{tad1}) can be expressed as
\begin{eqnarray}
\Sigma^{(a)} (t,\xi,\gamma,\delta )=-\left(\frac{2m_{0}^{\prime}\lambda\delta}{\pi^{2}}\right){\mathcal{R}}(1,t,\xi,\gamma,\delta),
\label{tad2}
\end{eqnarray}
where the dimensionless constant coupling  was defined as $\lambda = \lambda_0 m_{0}^{2}$ and the function ${\mathcal{R}}(1,t,\xi,\gamma,\delta)$ is given by
%%%%%%%%%%%%%%%%%%%%%%%%%%%%%%%%%%%%%%%%%%%%%%%%%%%%%%%%%%%
\begin{eqnarray}
{\mathcal{R}}(\nu,t,\xi,\gamma,\delta) &=& \sum_{s= \pm 1}^{}\sum_{\ell=0}^{+\infty}\left[\sum_{n_1=1}^{+\infty}(-1)^{n_1}\cosh(n_1\gamma/t)\left(\frac{2\pi n_1}{t\sqrt{\delta(2\ell+1-s)+1}}\right)^{\nu-1} \right. \nonumber \\
&&\times K_{\nu-1}\left(\frac{n_1}{t}\sqrt{\delta(2\ell+1-s)+1}\right) \nonumber \\ 
&&+\sum_{n_2=1}^{+\infty}(-1)^{n_2}\left(\frac{2\pi n_2}{\xi\sqrt{\delta(2\ell+1-s)+1}}\right)^{\nu-1} K_{\nu-1}\left(\frac{n_2}{\xi}\sqrt{\delta(2\ell+1-s)+1}\right)  \nonumber \\ 
&&+2\sum_{n_1,n_2 = 1}^{+\infty} (-1)^{n_1+n_2}\cosh(n_1\gamma/t)  \left(\frac{2\pi}{\sqrt{\delta(2\ell+1-s)+1}}\sqrt{\frac{n_1^{2}}{t^2}+\frac{n_2^2}{\xi^2}}\right)^{\nu-1} \nonumber \\
&&\left.\times K_{\nu-1}\left(\sqrt{\delta(2\ell+1-s)+1}\sqrt{\frac{n_1^{2}}{t^2}+\frac{n_2^2}{\xi^2}}\right)
\right].
\label{R}
\end{eqnarray}
\subsection{Shoestring diagram contribution}
An analogous calculation leads to the $2$-loop contribution
\begin{eqnarray}
\Sigma^{(b)}  (t,\xi,\gamma,\delta )= \left(\frac{8 m_{0}\eta \delta^{2}}{\pi^4}\right) \left[{\mathcal{R}}(1,t,\xi,\gamma,\delta)\right]^2,
\end{eqnarray}
where $\eta=\eta_0 m_{0}^{5}$.

The total correction to the mass parameter, at first order in the coupling constants, due to temperature, finite size, chemical potential, and magnetic field, for the system is, then, 
\begin{eqnarray}
m(t,\xi,\gamma,\delta) &=& m_{0}^{\prime}-\left(\frac{2m_{0}^{\prime}\lambda\delta }{\pi^{2}}\right){\mathcal{R}}(1,t,\xi,\gamma,\delta)+\left(\frac{8 m_{0}\eta \delta^{2}}{\pi^4}\right)\left[{\mathcal{R}}(1,t,\xi,\gamma,\delta)\right]^2. \nonumber \\
\label{massa}
\end{eqnarray}
In the next section, we solve numerically  Eq. (\ref{massa}) and describe the phase structure of the model.
\section{Phase Transition}
At criticality, we obtain different conditions to the first- and second-order phase transitions. For this reason, each case is presented separately. The modified Ginzburg--Landau free energy density is rewritten as 
\begin{eqnarray}
\frac{\mathcal{F}}{m_0^4} \equiv f(\varphi) = -\frac{m(t,\xi,\gamma,\delta)}{m_0}\varphi^2 - \frac{\lambda}{2}\varphi^4 + \frac{\eta}{3}\varphi^6,
\label{EL}
\end{eqnarray}
where we take  the rescaled order parameter $\phi: \varphi = \phi/ m_{0}^{3/2}$.
\subsection{First-Order Phase Transition}
To describe a first-order phase transition, we choose in Eqs. (\ref{massa}) and (\ref{EL}), $\eta > 0, \lambda > 0$. The Ginzburg--Landau free-energy density has three minima, one of them at the origin, $\varphi = 0$, the  other two symmetrically located with respect to the origin. Also, we have two local maxima, symmetrically located with respect to $\varphi = 0$. The criticality condition for a first-order phase transition, is obtained by noticing that when we are at criticality, the three minima the Ginzburg--Landau free-energy density, lie on the same horizontal line, which can be taken as the horizontal axis\cite{livro,Lebellac}.  
Mathematically, this can be represented by the conditions $f=d f /d\varphi=0$. Then using Eq.~(\ref{EL}) it is easily  shown that the corrected mass, at criticality, is given by 
\begin{eqnarray}
m(t_c,\xi,\gamma,\delta) = - \left(\frac{3 \lambda m_0}{16\eta}\right). 
\label{critica1}
\end{eqnarray}
Eq. (\ref{massa})  is to be replaced into Eq.~(\ref{critica1}) and we have to solve a transcendental equation for the critical temperature, $t_c$, in terms of the other parameters.
In  Fig.~$2$, we see the magnetic catalysis effect, i.e., the magnetic field contributes to decrease the critical temperature of the system (all figures in color in the online version). 
\begin{center}
\includegraphics[scale=0.7]{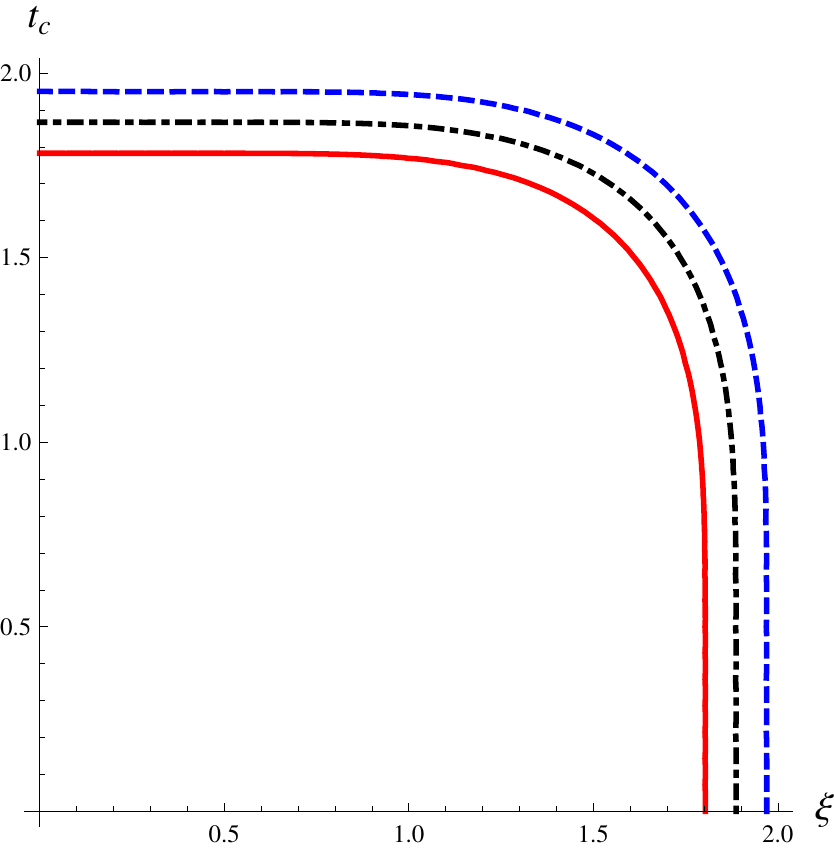}
\label{Fig2}
\end{center}
{Fig.~2 - \it{Critical temperature as a function of the inverse length of the system. For the three curves, we have fixed $\gamma = 0.5, \ \lambda = 2.0 \ $ and $\eta = 1.0$. Also, we have used $\delta = 1.5$ (dashed curve); $\delta = 3.0$ (dot-dashed curve); $\delta = 4.5$ (continuous curve).}}

We also see that as the field increases, $\xi_{max}$ decreases. In the other words, as the magnetic field becomes stronger, the minimal thickness of the system becomes larger, to ensure that the transition will happen.

From Fig.~$3$, we can see the small influence that the chemical potential has on the model. Furthermore, the minimal size of the system, $\xi_{max}$, seems to be independent of the chemical potential.
\begin{center}
\includegraphics[scale=0.7]{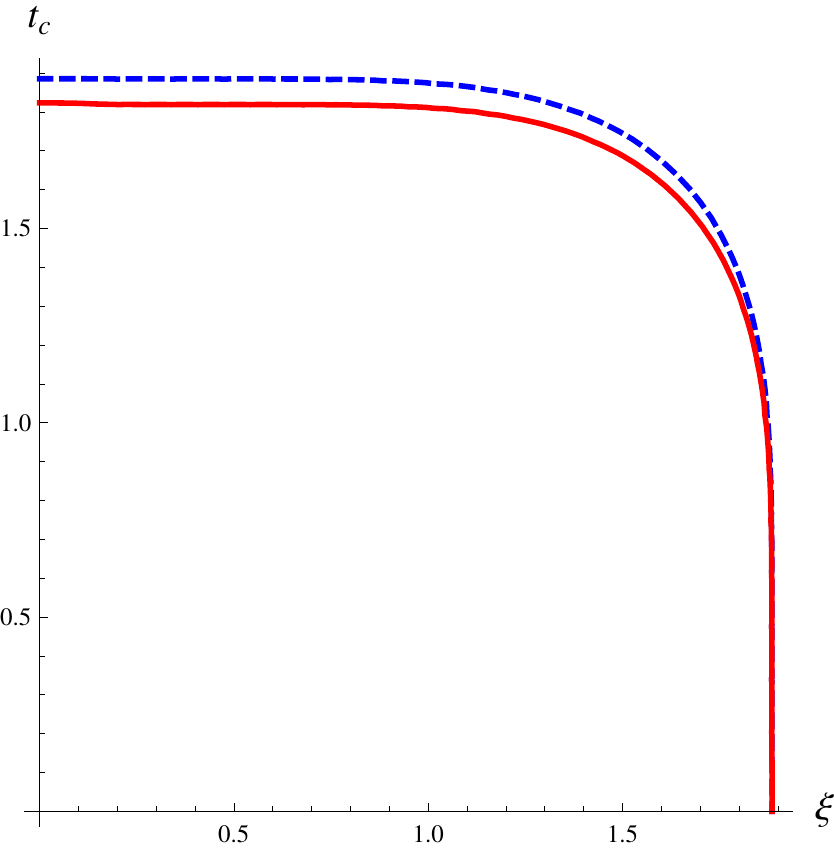}
\label{Fig3}
\end{center}
{Fig.~3 - \it{Influence of the chemical potential on the model. In the two curves, we have fixed $ \lambda = 2.0  $; \ $\eta = 1.0$ and $\delta = 3.0$. We have used $\gamma = 0.0$ (dashed curve); $\gamma = 0.75$ (full curve).}}

From Eqs. (\ref{massa}) and (\ref{R}), we have seen that the product ${\mathcal{P}}\equiv\cosh\left(\gamma n_1/t\right)
\times K_{0}[(n_1/t)\sqrt{\delta(2\ell+1-s)+1}]$ comes up. Since at small temperatures we can use the asymptotic formula $K_{\nu}(z)=\sqrt{\pi/2}\left[\exp(-z)\right]/\sqrt{z}$, valid when $|z|\rightarrow \infty$, $|\arg z| <\pi$, $Re \ {\nu}>-1/2$ and imposing the convergence condition $\gamma\leq\sqrt{\delta(2\ell+1-s)+1}$, we get that the product ${\mathcal{P}}$ goes to zero at the limit $t_c \rightarrow 0$. Therefore,  Eq.(\ref{massa}) is independent of $\gamma$ at $t_c \approx 0$ as shown in Fig.~$3$. A similar result was found in \cite{Emerson}.

In Fig.~$4$, we find two values of critical temperature that satisfy the condition $m(t_c,\xi,\gamma,\delta)/m_0 = - 3 \lambda /16\eta$: the smaller one, $t_{c1}$ (from Fig.~$4$, in the bulk form, i.e., $\xi \approx 0$ or $L \rightarrow 0$, its value is $t_{c1} \approx 0.76$) and the larger one, $t_{c2}$ (again from Fig.~$4$ and in the bulk form, $t_{c2} \approx 1.71$). This phenomenon is known as {\it{inverse symmetry breaking}}.
\begin{center}
\includegraphics[scale=0.7]{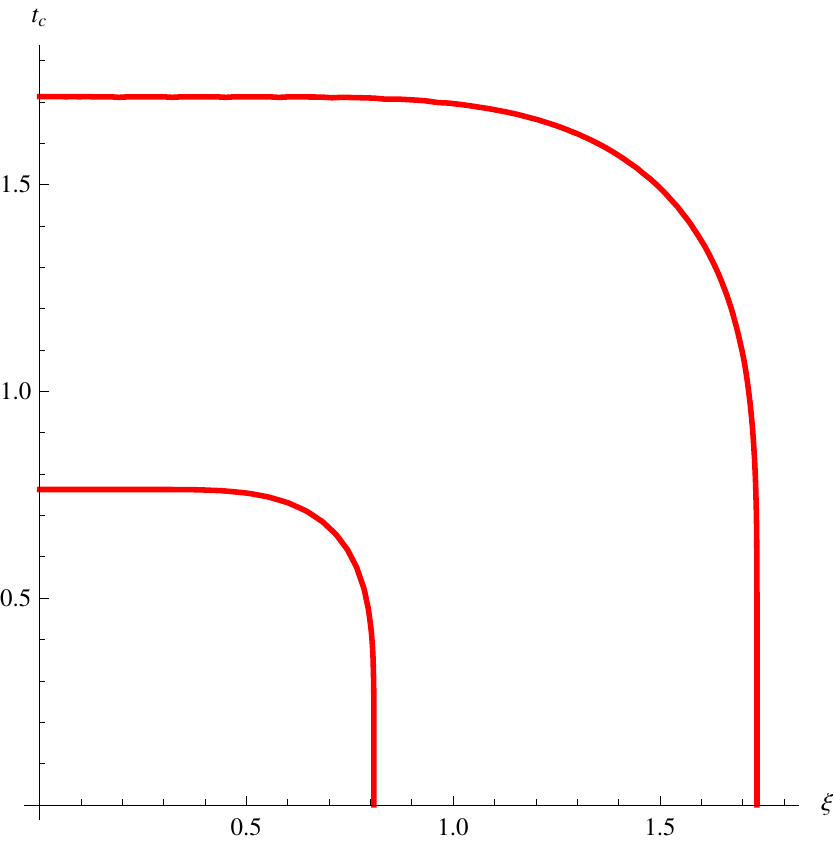}
\label{Fig4}
\end{center}
{Fig.~4 - \it{Inverse symmetry breaking phenomenon. There are two regions in which a phase transition occurs for the same values of $(\gamma,\delta, \lambda, \eta)$. We have fixed $\gamma = 0.5, \ \lambda = 4.0 \ $ and $\eta = 2.0$. We have used $\delta = 1.5$.}}

At temperatures lower than $t_{c1}$, we have that the inner region defined by $t_{c1}$ is the region of symmetry breaking. In the region between the two curves of the Fig.~$4$, i.e., with temperatures in the range $t_{c1} < t < t_{c2}$, there is a disordered phase. But this disordered phase does not persist. Indeed, at temperatures $t>t_{c2}$, we get again the ordered phase. Thus, to temperatures above $t_{c2}$, we also have a symmetry breaking region. As a consequence of the presence of these two critical temperatures, we have two minimal thicknesses of the system that support the transition.    

In  Figs.~$5$ and $6$, we show the behaviour of the Ginzburg--Landau free energy of the system in bulk form. Notice  that in Fig.~$5$, we have the usual behaviour of a first-order phase transition, i.e., we pass from the symmetric phase to the broken symmetry phase as the temperature decreases. On the other hand, we have the opposite behaviour in Fig.~$6$.
\begin{center}
\includegraphics[scale=0.8]{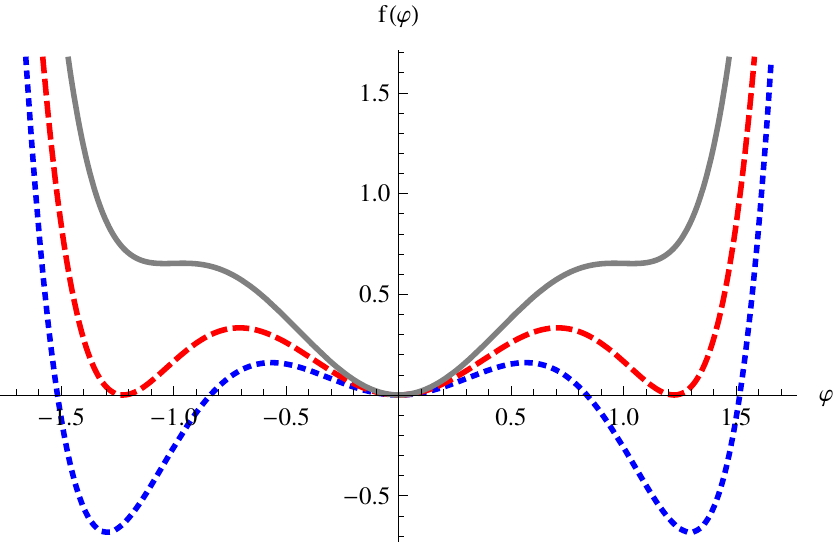}
\label{Fig5}
\end{center}
{Fig.~5 - \it{Free-energy density in the bulk region (inner part of Fig.~$4$). We observe a standard symmetry breaking. We have fixed $\gamma = 0.5, \ \lambda = 4.0 \ $ and $\eta = 2.0$. We have used $\delta = 1.5$. In the full curve, we have $t=1.4$. Criticality is achieved at $t_c \approx 0.76$ (dashed curve). Below $t_c$, we have the broken symmetry region, as usual (dotted curve, where $t = 0.3$).}}
\begin{center}
\includegraphics[scale=0.8]{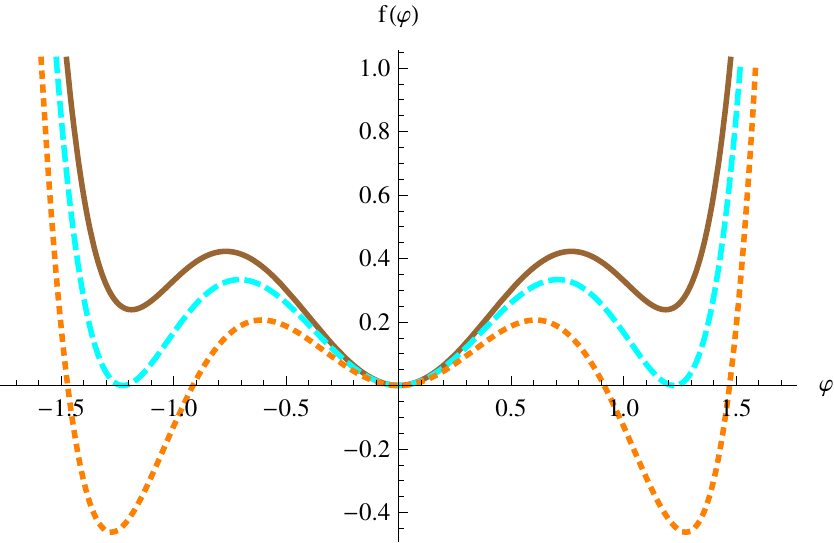}
\label{Fig6}
\end{center}
{Fig.~6 - \it{Free-energy density in the bulk (but now we are in external part of Fig.~$4$). We have an inverse symmetry breaking. We have fixed $\gamma = 0.5, \ \lambda = 4.0 \ $ and $\eta = 2.0$. We have used $\delta = 1.5$. In the full curve, we have $t=1.65$. Criticality is achieved at $t_c \approx 1.71$ (dashed curve). Even above $t_c$, we have a broken symmetry region (dotted curve, we fixed $t = 1.80$).}}
\subsection{Second-Order Phase Transition}
To describe a second-order phase transition, we have to fix $\eta = 0$ and choose $\lambda < 0$ in  Eqs. (\ref{massa}) e (\ref{EL}). The criticality condition is $m(t_c,\xi,\gamma,\delta) = 0$, as in the Ginzburg-Landau model. In  Fig.~$7$ we observe that the catalysis effect occurs again. 
\begin{center}
\includegraphics[scale=0.5]{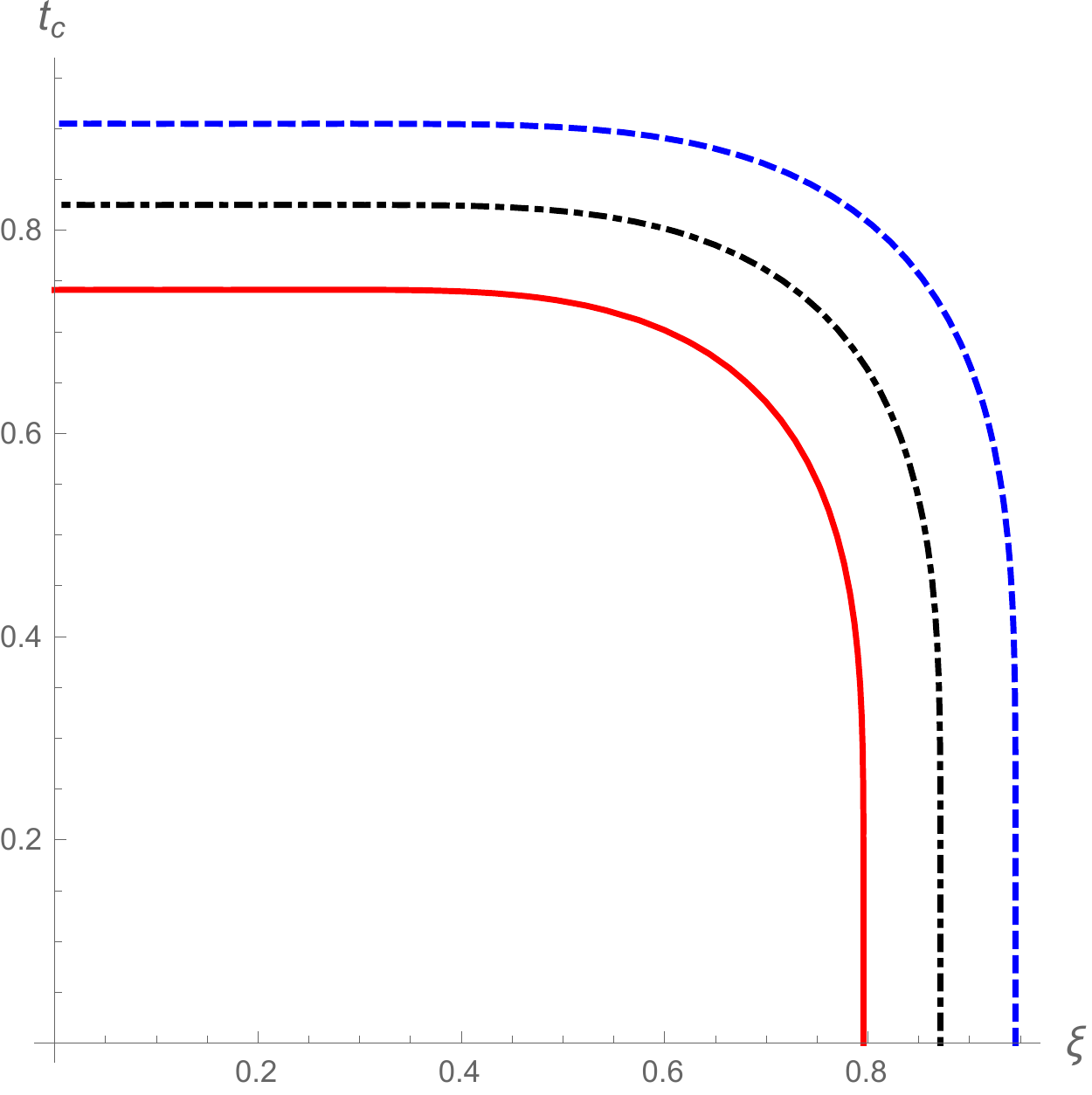}
\label{Fig7}
\end{center}
{Fig.~7 - \it{Critical temperature as a function of the inverse length. In the three curves, we have fixed $\gamma = 0.5, \ \lambda = 4.0 \ . $ We have used $\delta = 2.5$ (dashed curve); $\delta = 3.5$ (dot-dashed curve); $\delta = 4.5$ (full curve).}}

In the same way as in the case of the first-order phase transition, we notice a minimal size of the system below which the second-order phase transition disappears.

In Fig.~$8$, we observe a strong dependence of the coupling constant on the critical temperature.
\begin{center}
\includegraphics[scale=0.75]{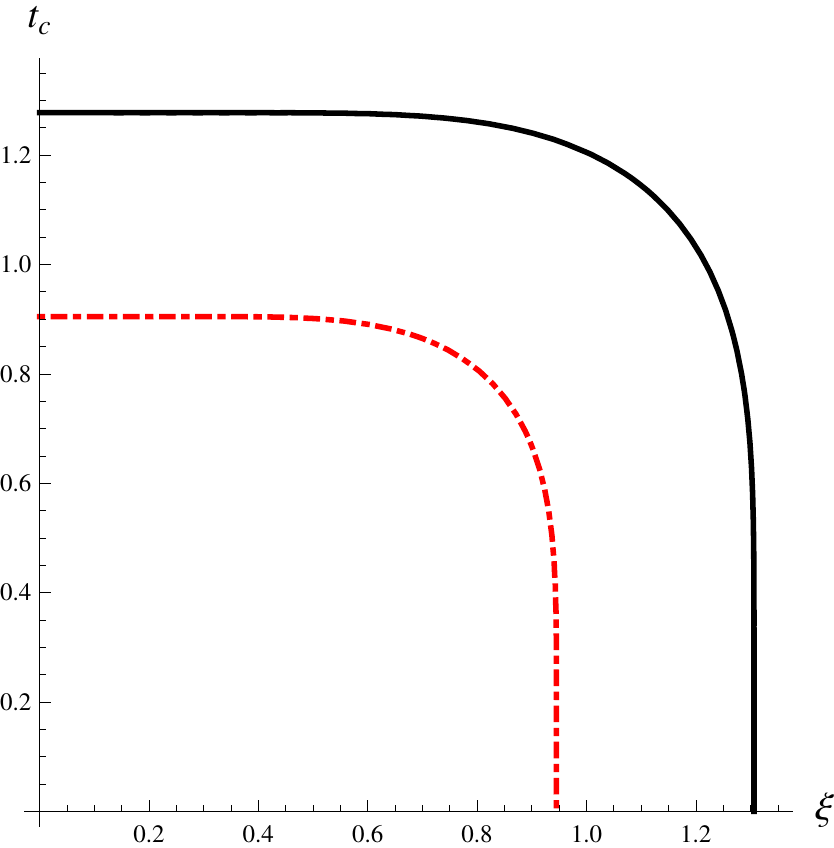}
\end{center}
{Fig.~8 - {\it{Critical temperature as a function of the inverse length for high values of the coupling constant. We have taken $\gamma = 0.5, \ \delta = 2.5 \ . $ We have used $\lambda = 2.0$ (full curve); $\lambda = 4.0$ (dot-dashed curve).}}}

In Fig.~$9$, we show the critical temperature behaviour of the model taking into account the chemical potential.
\begin{center}
\includegraphics[scale=0.75]{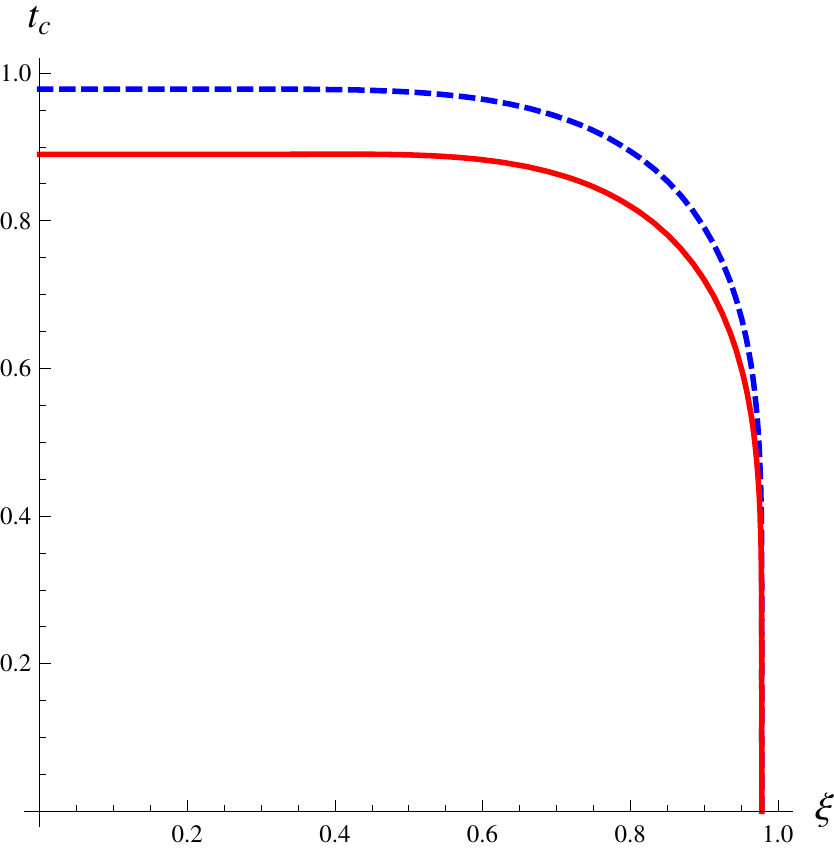}
\end{center}
{Fig.~9 - {\it{Critical temperature as a function of the inverse length for two values of the chemical potential. We have taken $\lambda = \delta= 2.0 $. We have fixed $\gamma = 0.0$ (dashed curve); \ $\gamma = 0.75$ (continuous curve).}}}

In Fig.~$10$, we show the Ginzburg--Landau free-energy density for the bulk region of Fig.~$7$.
\begin{center}
\includegraphics[scale=0.85]{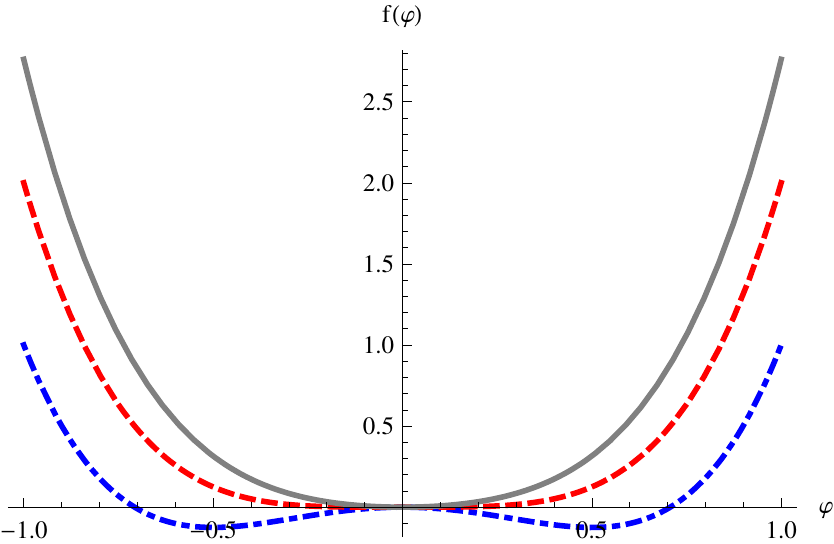}
\end{center}
{Fig.~10 - {\it{Free energy density for the second-order phase transition. We take $\delta= 2.5 $, $\lambda = 4.0$  and $\gamma = 0.5$. For the full curve, we have $t = 1.2$. The critical temperature occurs at $t_{c} \approx 0.9$ (dashed curve). In the dot-dashed curve, we have $t = 0.01$ \ .}}}

To get the critical temperature we have to numerically solve the transcendental equation $m(t_c,0,\gamma,\delta) = 0$, i.e., in the bulk region. We obtain that at temperatures above $t_c \approx 0.9$, we have a symmetric phase. On the other hand, at temperatures below $t_c \approx 0.9$, we have symmetry breaking. 
\section{Comments and Conclusions}
In this paper we present a model  to describe in a unified way first- and second-order phase transitions for a fermionic self-interacting system. Using results  from  quantum field theory on a toroidal topology and a one-component massive extended GN model, we have done an investigation about chemical potential, finite-size and magnetic effects for both transitions. In the first-order phase transition, as well as in the second-order one, we see that the system must have a minimal size, $L_{min}$, corresponding to a reduced
inverse size, $\xi _{max}$. There is no transition below this
minimal thickness. This is a reasonable result, since  our system corresponds to a heated film of thickness $L$ along the $z$-direction. Notice that the limiting situation $L = 0$ is not well defined in the formalism. We can see that the effect of the magnetic field is to lower the critical temperature, the so-called magnetic catalysis. We notice that the minimal size is independent of the chemical potential and that the
strength of the couplings have a significant influence on the phase transition of the system. Namely, for higher values of the couplings, we obtain larger minimal sizes. In particular, for the first-order transition, higher values of $\lambda $ and $\eta$ lead to two minimal sizes and two regions under transition. This phenomenon in called inverse symmetry breaking which is already discussed in the literature, for instance, in \cite{paperfermionico,Weinberg, Bimonte,Ramos,Hong,Caldas,Ramos1,Hoa,Sakamoto}. From Fig.~$4$, we can see that there exists an ordered phase at greater thicknesses and/or low temperatures (internal part of Fig.~$4$, as usual). Also, there is a transition for small thicknesses and/or high temperatures (external part of Fig.~$4$. In the region between the curves in Fig.~$4$, there is a disordered phase. In nature, the inverse symmetry breaking phenomenon has been observed in some composed systems, like the {\it{Rochelle Salt}}. This salt has an orthorhombic structure (ordered phase) at temperatures below $t_{1}\approx -18{}^{\circ}C$ and above $t_{2}\approx 24{}^{\circ}C$. For temperatures in the range $t_{1} < t < t_{2}$, the Rochelle Salt has a monoclinic unit cell (disordered phase). In  Refs.~\cite{Borut,Israelenses} we find other systems that suffer the same effect;, even water, under special pressure conditions, can present inverse symmetry breaking. We have quantitatively studied this behaviour in the model we have considered, for the system in bulk form, in Figs.~$5$ and $6$. We have found that both transitions suffer little influence of the chemical potential. In fact, we have shown explicitly that near $t_{c} \approx 0$, there is no dependence of  the model on the chemical potential,  as shown in Figs.~$3$ and $9$. In future developments, we would like to investigate whether these results hold for other backgrounds.
\section*{Acknowledgments} 
We thank financial support from CNPq (Brazilian agency). EBSC
also thanks the CAPES/PRODOUTORAL/UNIFESSPA program.
%%%%%%%%%%%%%%%%%%%%%%%%%%%%%%%%%%%%%%%%%%%%%%%%%%%%%%%%%%%%%%%%%%%%%%%%%%%%%%%%%%%%%%%%%%%%%%%%%%%%%%%%%%%%%%%%%%%%%%%%%%%%%%%%%%%%%%%%%%%%%%%%%%%%%%%%%%%%%%%%%%%%%%%%%%%%%%%%%%%%%%
%

\end{document}